\documentclass[prl,twocolumn,superscriptaddress,showpacs,amsmath,amssymb]{revtex4-1}

\usepackage{graphicx}

\begin{document}
\title{Observation 
of time quasicrystal and its transition to superfluid time crystal}

\author{S.~Autti}
\affiliation{Low Temperature Laboratory, Department of Applied Physics, Aalto University, POB 15100, FI-00076 AALTO, Finland. Email: samuli.autti@aalto.fi}
\affiliation{Department of Physics, Lancaster University, Lancaster LA1 4YB, UK}

\author{V.B.~Eltsov}
\affiliation{Low Temperature Laboratory, Department of Applied Physics, Aalto University, POB 15100, FI-00076 AALTO, Finland. Email: samuli.autti@aalto.fi}
\author{G.E.~Volovik}
\affiliation{Low Temperature Laboratory, Department of Applied Physics, Aalto University, POB 15100, FI-00076 AALTO, Finland. Email: samuli.autti@aalto.fi}\affiliation{L.D. Landau Institute for Theoretical Physics, Moscow, Russia}

\date{\today}
\begin{abstract}

We report experimental realization of a quantum time quasicrystal, and its transformation to a quantum time crystal. We study Bose-Einstein condensation of magnons, associated with coherent spin precession, created in a flexible trap in superfluid $^3$He-B. Under a periodic drive with an oscillating magnetic field, the coherent spin precession is stabilized at a frequency smaller than that of the drive, demonstrating spontaneous breaking of discrete time translation symmetry. The induced precession frequency is incommensurate with the drive, and hence the obtained state is a time quasicrystal. When the drive is turned off, the  
self-sustained coherent precession lives a macroscopically-long time, now representing a time crystal with broken symmetry with respect to continuous time translations. Additionally, the magnon condensate manifests spin superfluidity, justifying calling the obtained state a time supersolid or a time super-crystal.
 \end{abstract}



\maketitle

Originally time crystals were suggested as class a quantum systems for which time translation symmetry is
spontaneously broken in the ground state, so that the time-periodic motion of the background constitutes its lowest energy state \cite{Wilczek2013}. It was quickly shown that the original idea cannot be realized with realistic assumptions \cite{Bruno2013a,Bruno2013b,Nozieres2013,Volovik2013,Watanabe2015}. This no-go theorem forces us to search for spontaneous time-translation symmetry breaking in a broader sense (see e.g.\ review in Ref.~\cite{TimeCrystalsReview2017}). One available direction is systems with off-diagonal long range order, experienced by superfluids, Bose gases, and magnon condensates\cite{Volovik2013,Svistunov2017}. In the grand canonical formalism, the order parameter of a Bose-Einstein condensate (BEC) --- the macroscopic wave function $\Psi$, which also describes conventional superfluidity --- oscillates periodically: $\Psi=\langle\hat a_0\rangle=|\Psi|e^{-i\mu t}$, where $\hat a_0$ is the particle annihilation operator and $\mu$ is chemical potential. Such a periodic time evolution can 
be observed experimentally provided the condensate is coupled to another condensate. If 
the system is strictly isolated, i.e., when number of atoms $N$ is strictly conserved, there is no reference frame with 
respect to which this time dependence can be detected. That is why for the external observer, the BEC looks like a fully stationary ground state. 

However, for example in the Grand Unification  extensions of Standard Model there is no conservation of the number of atoms $N$ due to proton decay \cite{Nath2007}.  Therefore, in principle, the oscillations of the macroscopic wave function of an atomic superfluid in its ground state could be identified experimentally and the no-go theorem avoided if we had enough time for such experiment, about $\tau_N \sim 10^{36}$ years. In general, any system with off-diagonal long range order can be characterized by two relaxation
times \cite{Volovik2013}. One is the life time $\tau_N$ of the corresponding  particles (quasiparticles).  The second one is the thermalization time, or  energy relaxation time $\tau_E$, during which the superfluid state of $N$ particles is formed. If $\tau_E\ll \tau_N$, the system relatively quickly relaxes to a minimal energy state with quasi-fixed $N$ (the superfluid state), and then slowly relaxes to the true equilibrium state with $\mu=0$. In the intermediate time $\tau_E\ll t\ll\tau_N$ the system has finite $\mu$, and thus becomes a time crystal. Note that in the limit of exact conservation of particle number mentioned above, $\tau_N\rightarrow \infty$, the exchange of particles between the system and the environment is lost and the time dependence of the condensate cannot be experimentally resolved.

\begin{figure}
\centerline{\includegraphics[width=1.0\linewidth]{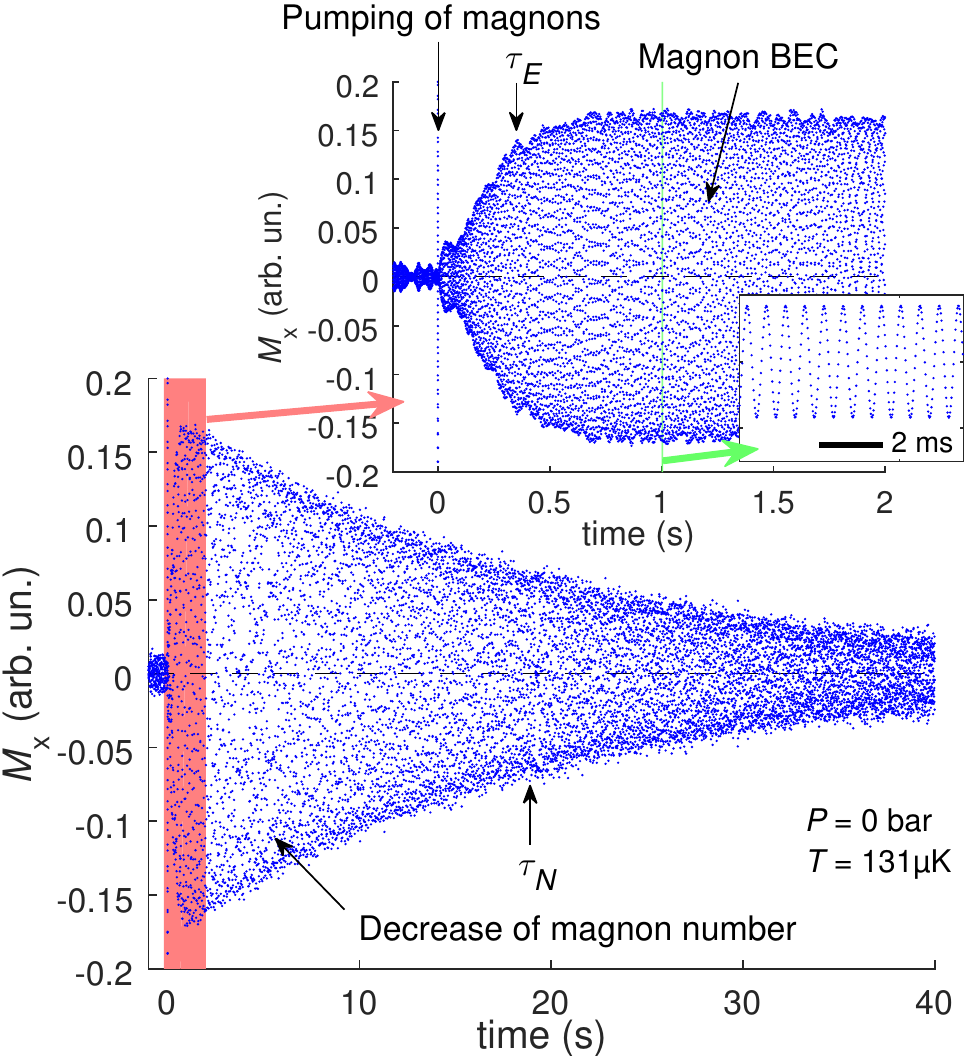}}
\caption{(Color online) Time super-crystal in superfluid $^3$He-B emerging from spontaneous coherence in freely precessing magnetization.
({\it Top}) The coherent precession of magnetization, $M_x+iM_y = \gamma S_\perp e^{i\omega t}$, is established with time constant $\tau_E$ after the excitation pulse at $t=0$. The signal is picked up by the NMR coils (Fig.~\ref{Cell}) and downconverted to lower frequency (with reference at 834\,kHz).  ({\it Bottom}) Since the magnetic relaxation in superfluid $^3$He is small, the number of magnons $N$ slowly decreases with time scale $\tau_N\gg \tau_E$. During relaxation the precession remains coherent and the state represents Bose-Einstein condensate of magnons until the number of magnons drops below the critical value (which is these experiments corresponds to signal below the noise level). 
}
\label{Precession}
\end{figure}

Bose-Einstein condensates of pumped quasiparticles, such as photons \cite{Klaers2010}, are in general a good example of systems with off-diagonal long-range order, where the condition $\tau_N\gg\tau_E$ is fulfilled. Time-crystals can be conveniently studied in experiments based on the magnon BEC states in superfluid phases of $^3$He, where the life time of quasiparticles (magnons) can reach minutes. Magnon BEC in $^3$He was first observed in the fully gapped topological B phase \cite{HPD1,HPD2}, then in the chiral Weyl superfluid A phase \cite{Aphase,CPA}, and recently in the polar phase with Dirac lines \cite{Autti2017}. Magnon number $N$ is not conserved, but the decay time $\tau_N\gg\tau_E$, see Fig. \ref{Precession}. For $t < \tau_N$ the magnon BEC corresponds to the minimum of energy at fixed $N$. The life time $\tau_N$ is long enough to observe the Bose-condensation and effects related to the spontaneously broken $U(1)$ symmetry, such as ac and dc Josephson effects, phase-slip 
processes, Nambu-Goldstone modes, etc \cite{BunkovVolovik2013}. Each magnon carries spin $-\hbar$, and the number of magnons is thus $N=(S-S_z)/\hbar$, where $\mathbf S$ is the total spin and the $\hat{\mathbf z}$ axis is directed along the applied magnetic field $\mathbf H$. 
The state with $|S_z| < S$ corresponds to the precessing macroscopic spin with transverse magnitude $S_\perp=\sqrt{S^2-S_z^2}$ and  the broken $U(1)$ symmetry within the condensate is equivalent to broken $SO(2)$ symmetry of spin rotation about the direction of $\mathbf H$. The magnon BEC is manifested by spontaneously formed coherent spin precession, and is described by the wave function $\langle\hat S^+\rangle= \sqrt{2S}\left<\hat a_0\right>=S_\perp e^{i\omega t}$, where $\hat S^+ = \hat S_x + i \hat S_y$. Here the role of the chemical potential is played by the global frequency $\omega$ of the precession. A characteristic feature of coherent precession is that this frequency is constant in space even for a spatially inhomogeneous magnon condensate in an inhomogeneous trapping potential.  

\begin{figure}
\centerline{\includegraphics[width=0.75\linewidth]{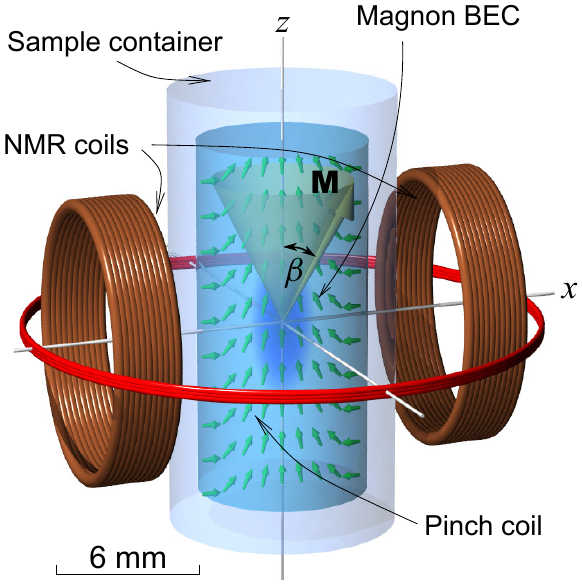}}
\caption{(Color online) Experimental setup: Sample container is made from quartz glass and filled with superfluid $^3$He-B. NMR coils create pumping field $\mathbf H_{\rm rf}$ and pick-up induction signal from coherently precessing magnetization $\mathbf M$ of the magnon BEC. The trapping potential for the condensate is created by a spatial distribution of the orbital anisotropy axis of $^3$He-B (small green arrows) and an axial minimum in the static magnetic field produced by a pinch coil.
}
\label{Cell}
\end{figure}

The coherently precessing spins are observed in NMR experiments via corresponding precession of magnetization $\mathbf M = \gamma \mathbf S$, where $\gamma$ is the gyromagnetic ratio (Fig. \ref{Cell}). Equilibrium non-zero magnetization $\mathbf M = \chi \mathbf H$ is created by an applied static magnetic field, $\chi$ being the magnetic susceptibility. The magnetic field defines the Larmor frequency $\omega_L= |\gamma|  H$. Then a transverse radio-frequency (rf) pulse, $\mathbf H_{\rm rf} = H_{\rm rf} \,\hat{\mathbf x}\, e^{i\omega_{\rm drive} t}$, is applied to deflect the spins by angle $\beta$ with resperct to ${\bf H}$. This corresponds to pumping $N=S(1-\cos\beta)/\hbar$ magnons to the sample. After the pulse, the signal picked up by the NMR coils rapidly decays due to dephasing of the precessing spins caused by inhomogeneity of the trapping potential. After time $\tau_E$, collective synchronization of the precessing spins takes place, leading to the formation of the magnon BEC with off-diagonal long 
range order. This process is the experimental signature of the time crystal: the system spontaneously chooses a coherent precession frequency, and one can directly observe the resulting periodic time evolution (Fig.~\ref{Precession}).

Note that the periodic phase-coherent precession emerges in the interacting many-body magnon system,
which experiences the spin superfluidity \cite{BunkovVolovik2013}. The spontanously broken $U(1)$ symmetry and the interaction between magnons in magnon BEC give rise to the Nambu-Goldstone modes, which can be identified with phonons of this time crystal. 
The frequency of the precession is determined by interactions (here of spin-orbit type, see below) and is robust to perturbations in the system. All this fits the presently adopted criterions of  "time crystal", which exhibits
typical hallmarks of spontaneous symmetry breaking, such as long-range order and soft modes \cite{HerculeanTasks}.
 The coherent precession can be also considered as macroscopic realization of the time crystal behavior of excited eigenstates \cite{Sacha2017}.

Also another direction to circumvent the no-go theorem has been suggested -- Floquet time crystals emerging under a drive \cite{Sacha2015,Khemani2016,Khemani2016b,Else2016,Else2017,Yao2017,ZhangHess2017,Choi2017,Gibney2017,Abanin2017,Ueda2018}. As distinct from the breaking of continuous time-translation \cite{Wilczek2013,Bruno2013a,Bruno2013b,Nozieres2013,Volovik2013}, here the discrete time symmetry $t  \rightarrow  t + T$ is spontaneously broken, $T$ being the period of the driving force. The system spontaneously acquires a larger period than that of the drive, $nT$, i.e. $\omega_{\rm coherent}=\omega_{\rm drive}/n$, where $\omega_{\rm drive} = 2\pi/T$. For example, the period may be doubled, $n=2$ \cite{Russomanno2017}.
In Ref. \cite{Zavjalov2016}, a parametric resonance was observed, in which the magnon BEC generates pairs of (longitudinal) Higgs modes with $\omega_{\rm Higgs} = \omega_{\rm drive}/2$, although no direct demonstration of doubling of the period of the response of the magnon BEC itself was provided. The breaking of discrete time symmetry may also result in the formation of \emph{time quasicrystals} \cite{Flicker2017,Potter2018,Giergiel2017,Hou2018,TongcangLi2012}, where the periodic drive gives rise to the quasiperioic motion with, say, two incommensurate periods. We now discuss an observation of a time quasicrystal in a magnon BEC obtained by applying a drive, and its evolution to a superfluid time crystal (time super-crystal) when the drive is switched off.

In our experiments, the magnon BEC is trapped in a potential well. The trap is formed by the combination of the spatial distribution of the orbital anisotropy axis of the Cooper pairs, called texture, and by a magnetic field minimum, created with a pinch coil (Fig.~\ref{Cell}). The potential well is harmonic, and the magnon condensate can be excited on several different energy levels in it, not only in the ground state. The  ground state can be simultaneously populated by relaxation from the chosen exited level, forming a system of two coexisting condensates \cite{Autti2012}. Similar off-resonant excitation of the coherent spin precession was first observed in Ref. \cite{Cousins1999}.
It requires an excitation at a higher frequency than the
 frequency of the coherent precession in the ground state, in partial analogy with lasers \cite{Fisher2003} in the sense that a coherent signal emerges from incoherent pumping.

\begin{figure}
\centerline{\includegraphics[width=0.95\linewidth]{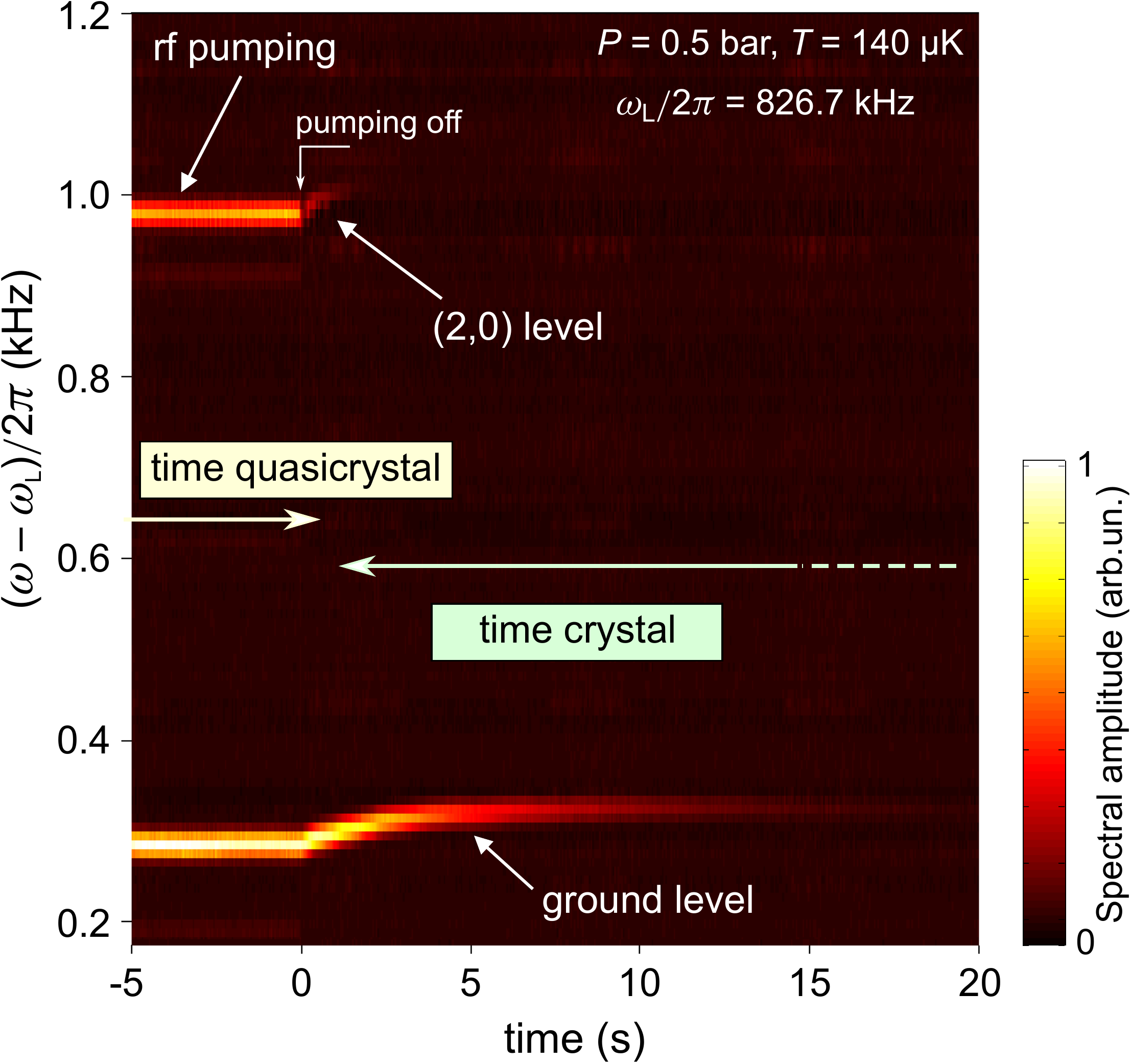}}
\caption{(Color online) Time crystal and time quasicrystal. The measured signal is analyzed using time-windowed Fourier transformation, revealing two distinct states of coherent precession seen as sharp peaks. Driving field $\textbf H_{\rm rf}$ with the frequency $\omega_{\rm drive}$, applied at $t < 0$, excites magnons on the second radial excited level in the harmonic trap. The majority of these magnons moves to the ground level, where they form the magnon BEC. Their precession is seen as the  signal at the smaller frequency, $\omega_{\rm coherent}<\omega_{\rm drive}$. The total signal is quasiperiodic in time, see Fig.~\ref{CMEfigure}. At $t=0$ the rf pumping is switched off and the (2,0) level quickly depopulates. Only the periodic signal from the precession of the ground-level magnon BEC remains, like in Fig.~\ref{Precession}. Its frequency is slowly increasing in time in the flexible trap, following the decay of the magnon number $N$. This emphasizes robust nature of breaking of continuous time 
translation symmetry and signifies the time crystal. 
}
\label{TQC}
\end{figure}

One important property of magnon condensates in the textural trap --- as compared with, say, atomic BECs in ultracold gases \cite{PitaevskiiStringari2003} ---  is that the trap is flexible. The trap is modified by the magnon BEC, which owing to the spin-orbit interaction repels the texture and extends the trap.
 As a result the energy levels in the trap depend on the magnon number $N$ in the condensate. This mechanism is similar to the formation of hadrons
in the so-called  MIT bag model in quantum chromodynamics (QCD), in which free quarks dig a box-like hole in the QCD vacuum \cite{MITBag1,MITBag2}. And indeed, in the limit $N\rightarrow \infty$ the harmonic trap transforms to the box \cite{Autti2012,Autti2012b}.
The flexible trap provides an effective interaction between the magnons. In atomic BECs interactions also lead to dependence of the chemical potential on the number of particles, but the functional dependence on $N$ is different. The eigenstates in the magnon trap determine possible frequencies of the coherent precession.
The dependence of the precession frequency on $N$ is seen in Fig. \ref{TQC} at $t>0$: during decay of the magnon BEC its ground-state energy level increases as the trap shrinks in size and eventually reaches the undisturbed harmonic shape.

\begin{figure}
\centerline{\includegraphics[width=0.95\linewidth]{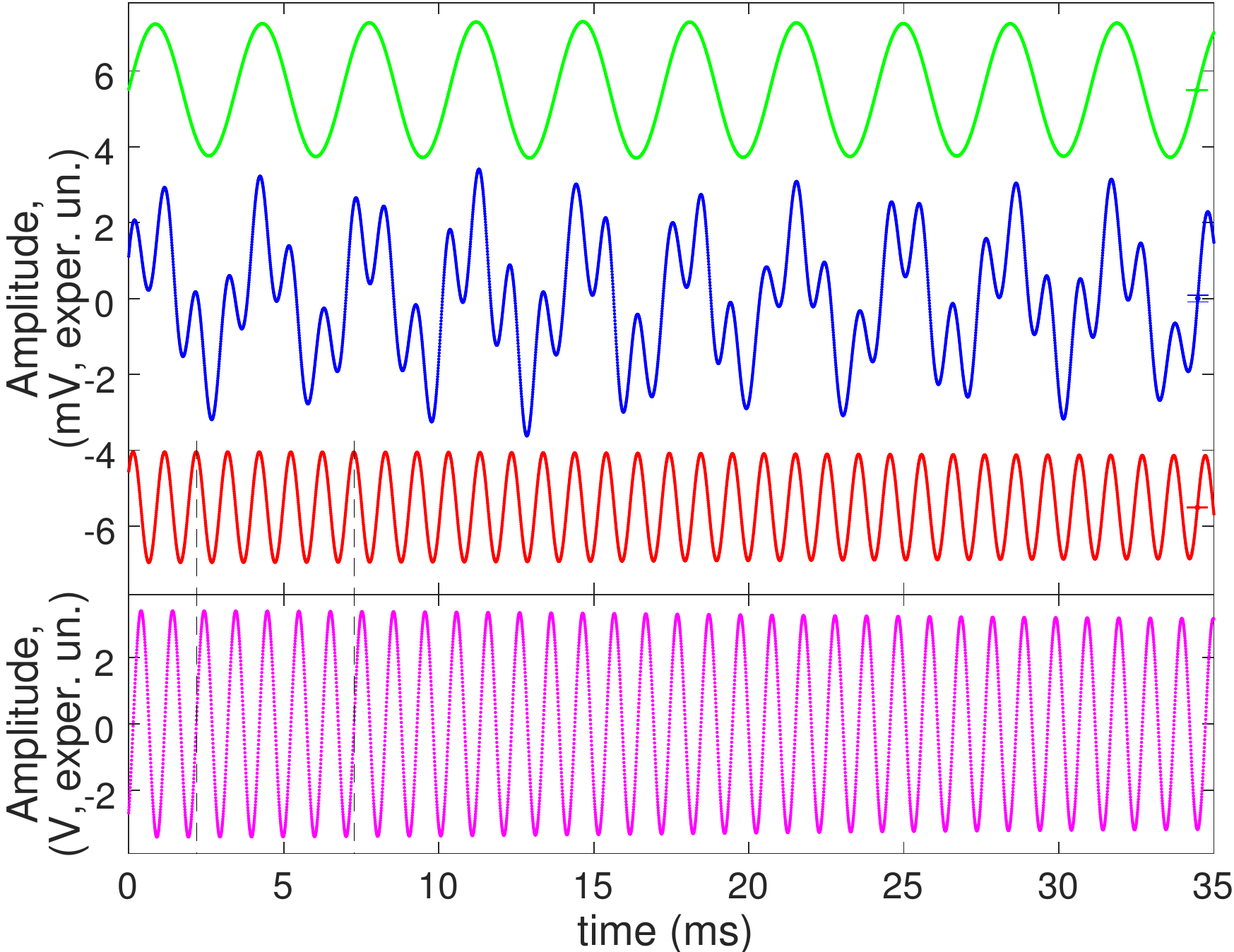}}
\caption{(Color online) Quasi-periodic signal from time quasicrystal at $t<0$  in Fig.~\ref{TQC} is shown in time domain. The section plotted here begins at $t=-4.97$\,s. The amplified signal from NMR coils is downconverted to lower frequency using Larmor frequency as a reference and digitized at 44\,kHz rate. From the total signal the drive (lower plot) is subtracted, as calibrated in a separate measurement without magnon BEC. The quasi-periodic signal (middle blue line in the upper plot) is then revealed. It is further splitted using band-pass filters to the signal from the ground-level magnon BEC (upper green line) at $(\omega_{\rm coherent} -\omega_{\rm L})/2\pi\approx290$Hz and signal from the (2,0) level at $(\omega_{\rm drive} -\omega_{\rm L})/2\pi\approx980$Hz (lower red line). These lines have been shifted vertically for clarity. Error bars on the right denote measurement noise amplitude in the respective frequency bands. Oscillations at (2,0) level proceed at exactly the drive frequency as shown by the dashed vertical lines. The frequency locking occurs automatically by self-tuning of the 
steady-state number of magnons \cite{Autti2012}. 
}
\label{CMEfigure}
\end{figure}

In Fig. \ref{TQC}  the frequency $\omega_{\rm drive}$ of the driving rf field corresponds to that of the second radial axially symmetric excited state in the harmonic trap (level (2,0)). The drive pumps magnons to this level forming magnon condensate there. One can see additional oscillations spontaneously  generated at a lower frequency $\omega_{\rm coherent}<\omega_{\rm drive}$, which corresponds to the BEC forming at the ground state in the trap by magnons coming from the exited state. Note that the frequencies of the ground and exited states can be tuned  by changing the magnetic field magnitude and, independently, the spacing of the states by the magnetic field profile. The spacing further depends on the number of magnons in the trap, and one can choose the frequencies to be incommensurate. This demonstrates that the discrete time symmetry, $t  \rightarrow  t + T$, of the drive is spontaneously broken leaving a state composed of precesison at two incommensurate frequencies $\omega_{\rm coherent}$ and $\omega_{\rm drive}$. We emphasize that the two states coexist in the same trapping potential and hence occupy the same volume. The observed state is therefore a time quasicrystal (the term time quasicrystal was introduced in Ref. \cite{Yin2012}). The measured signal with oscillations at two well-resolved frequencies is shown in Fig.~\ref{CMEfigure}.

After the pumping is stopped  (at $t=0$ in Fig. \ref{TQC}), the excited-state condensate rapidly decays. What is left is the condensate in the ground state, whose frequency slowly increases with time following the decay of the magnon number $N$. This state is a time crystal, as discussed above. That is, the driven time quasicrystal with broken discrete time translation symmetry transforms to the time crystal with broken continuous time symmetry.

In conclusion, in a single experiment we have observed both types of time crystals discussed in the literature. These are states with broken continuous and discrete time translation symmetries. They are found in coherent spin dynamics of superfluid $^3$He-B, interpreted using the language of magnon Bose-Einstein condensation in a flexible trap provided by the $^3$He-B order parameter distribution. The discrete time translation symmetry breaking takes place under an applied rf drive. The magnon condensation is then manifested by coherent spin precession at a frequency smaller than the drive. The induced precession frequency is incommensurate with the drive, giving rise to a time quasicrystal with the discrete time-translation symmetry being broken. When the drive is turned off, the
self-sustained coherent precession lives for a long time, while the number of magnons decays only slowly. 
This is a time crystal. Both the time crystal and the time quasicrystal are formed in the 
topological superfluid $^3$He-B \cite{Mizushima2016,MizushimaSauls2018} and possess spin superfluidity. Therefore these states can be called time supersolids.

We thank Jaakko Nissinen for stimulating discussions. This work has been supported by the European Research Council (ERC) under the European Union's Horizon 2020 research and innovation programme (Grant Agreement No. 694248). The work was carried out in the Low Temperature Laboratory, which is part of the OtaNano research infrastructure of Aalto University. S. Autti acknowledges financial support from the Jenny and Antti Wihuri foundation.

\end{document}